\documentclass[12pt]{article}
\usepackage{psfig,epsf,amssymb,amsmath,latexsym}
\title{Lifshitz tails for the $1D$ Bernoulli-Anderson model}

\author{H. Schulz-Baldes
\\
\\
{\small Fachbereich Mathematik, Technische Universit{\"a}t Berlin, 
10623 Berlin, Germany}
\\
\\
}

\date{ }

\newcommand{\RR}{{\mathbb R}}

\newcommand{\ZZ}{{\mathbb Z}}

\newcommand{\PP}{{\bf P}}
\newcommand{\EE}{{\bf E}}

\newcommand{\Ss}{{\cal S}}
\newcommand{\Oo}{{\cal O}}
\newcommand{\Tr}{\mbox{\rm Tr}}

\newcommand{\Nn}{{\cal N}}

\begin{document}

\maketitle

%%%%%%%%%%%%%%%%%%%%%%%%%%%%%%%%%%%%%%%%%%%%%%%%%%%%
\begin{abstract}
By using the adequate modified Pr\"ufer variables,
precise upper and
lower bounds on the density of states in the (internal) 
Lifshitz tails are proven
for a $1D$ Anderson model with bounded potential.
\end{abstract}
%%%%%%%%%%%%%%%%%%

\vspace{.5cm}

There have been numerous rigorous works about Lifshitz tails for the
$1D$-Anderson model with bounded potentials (see \cite{KW} for a
collection of references). The aim of
this note is to give a simple proof by passing to the normal form of
the transfer matrix at a band edge and then using the adequate
Pr\"ufer variables in this regime. This allows to obtain quite
precise estimates on the integrated density of states (IDS).
The paper concludes with a brief outlook on how and why similar
techniques lead to perturbative
results about the IDS and the Lyapunov exponent in between the
Lifshitz tails and the center of the band, a regime that 
will be studied in more detail elsewhere. 

%%%%%%%%%%%%%%%%%%%%%%%%%%%%%%%%%%%%%%%%%%
\section{Result}
\label{sec-result}

Let $(H_\sigma)_{\sigma\in\Sigma}$ be a family of $L$-periodic real
Jacobi matrices on $\ell^2(\ZZ)$. Each $H_\sigma$
is specified by $2L$ real numbers $t_\sigma(l)\geq 0$ and
$v_\sigma(l)$ where $l=1,\ldots,L$ such that for  
any state $\psi\in\ell^2(\ZZ)$

$$
(H_\sigma\psi)(n)\;=\;
-t_\sigma(n+1)\psi(n+1)+v_\sigma(n)\psi(n)-t_\sigma(n)\psi(n-1)
\mbox{ . }
$$

\noindent The eigenvalue equation $H_\sigma\psi=E\psi$ for $E\in\RR$
can conveniently be rewritten using the transfer matrices
$T^E_\sigma\in SL(\RR,2)$ over one period:

$$
\left(\begin{array}{c} t_\sigma(L)\,\psi(L) \\ \psi(L-1)
\end{array} \right)
\;=\;
T^E_\sigma\;
\left(\begin{array}{c}t_\sigma(1)\, \psi(1) \\ \psi(0)
\end{array} \right)
\mbox{ . }
$$

\vspace{.2cm}

The set $\Sigma$ is supposed to be finite here and on it be
given a probability measure ${\bf p}=\sum_\sigma
p_\sigma\delta_\sigma$. Of course, $p_\sigma\geq 0$ and
$\sum_\sigma p_\sigma=1$. The Tychonov space
$\Omega=\Sigma^{\times \ZZ}\times\{1,\ldots L\}$ 
is furnished with the probability
measure $\PP={\bf p}^{\otimes \ZZ}\otimes
\frac{1}{L}\sum_l\delta_l$. 
This measure is of the so-called Bernoulli
type and it is invariant and ergodic w.r.t. the natural translation
action of $\ZZ$ on $\Omega$. Associated to each configuration
$\omega=((\sigma_n)_{n\in\ZZ},l)\in\Omega$ is a Hamiltonian $H_\omega$
obtained by juxtaposition of the periodic blocs according to the
configuration $\omega$. More precisely, if $n=mL+l+k$ with $k=1,\ldots,L$, then

$$
(H_\omega\psi)(n)\;=\;
-t_{\sigma_m}(k+1)\psi(n+1)+v_{\sigma_m}(k)\psi(n)-t_{\sigma_m}(k)\psi(n-1)
\mbox{ , }
$$

\noindent with the convention that
$t_{\sigma_m}(L+1)=t_{\sigma_{m+1}}(1)$.  
By construction, $(H_\omega)_{\omega\in\Omega}$ is a strongly
continuous operator family for which the covariance relation
w.r.t. the translations holds \cite{PF,Bel}. 
Note that, even in the trivial case where ${\bf p}$ is supported by
only one point $\sigma\in\Sigma$, one has a covariant operator family
given by the $L$ shifts of the $L$-periodic operator $H_\sigma$.
The spectrum of any such covariant family is
$\PP$-almost surely constant. Using approximate eigenfunctions, one
easily verifies $\mbox{spec}(H_\omega)\subset
\bigcup_{\sigma,\,p_\sigma\neq 0}
\mbox{spec}(H_\sigma)$. Equality instead of inclusion holds if
$H_\sigma$ is the sum of a periodic background operator and a random
potential \cite{PF}. In this situation, a band edge of the random
operator is a band edge also for one of the periodic operators.

\vspace{.2cm}

Two fundemmental objects associated to a covariant 
family of $1D$ operators
are the IDS $\Nn$ and the Lyapunov exponent
\cite{PF}, the focus here being only on the former. One of the
equivalent definitions of the Stieltjes function $\Nn$ is formula
(\ref{eq-IDS}) in Section \ref{sec-setup}. Its support is  
$\mbox{spec}(H_\omega)$. If ${\bf p}=\delta_\sigma$, 
then it is straight-forward to calculate the
associated IDS, denoted $\Nn_\sigma$, by means of 
Bloch-Floquet theory.

\vspace{.2cm}

Let now $E_\nu\in\RR$ be a boundary point of $\mbox{spec}(H_\nu)$ and 
$\mbox{spec}(H_\omega)$. Hence $E_\nu$ can be either the bottom or the
top of the spectrum or 
the boundary point of an {\it internal} spectral gap. Within any gap, the
IDS $\Nn(E_\nu)$ is constant and therefore it
is natural to study

$$
\delta\Nn(\epsilon)
\;=\;
\pm\,\Nn(E_\nu\pm\epsilon)\,\mp\,\Nn(E_\nu)
\mbox{ , }
$$

\noindent where the upper signs are chosen if $E_\nu$ is a lower band
edge and the lower signs for an upper band edge.
As above, one also introduces $\delta\Nn_\nu(\epsilon)$. 
Close to $E_\nu$, the IDS $\Nn$ is very small and its scaling is universal and
called
a Lifshitz tail, honoring the original contribution of L. Pastur's
teacher. 

\vspace{.3cm}

%%%%%%%%%%%%%%%%%%%%%%%%%%%%%%%%%%%%%%%%%%%%%%
\noindent {\bf Theorem}
{\it
Let $E_\nu$ be a band edge of $H_\nu$ and $H_\omega$, but not of $H_\sigma$  
if $\sigma\neq\nu$.
Suppose moreover that the eigenvector of
$T^{E_\nu}_\nu$ is not an eigenvector of $T^{E_\nu}_\sigma$ for any
other $\sigma\neq\nu$. Then there exist constants $C>0$ and 
$\epsilon_0>0$ such that for
all $\epsilon\in [0,\epsilon_0]$

\begin{equation}
\label{eq-main}
\frac{1}{2}\;\delta\Nn_\nu(\epsilon)\;
p_\nu^{\frac{1}{L\,\delta\Nn_\nu(\epsilon)}+1}
\;\leq\;
\delta\Nn(\epsilon)
\;\leq\;
C\,\delta\Nn_\nu(\epsilon)\;
p_\nu^{\frac{1}{L\,\delta\Nn_\nu(\epsilon)}}
\mbox{ . }
\end{equation}
}
%%%%%%%%%%%%%%%%%%%%%%%%%%%%%%%%%%%%%%%%%%%%%%

\vspace{.3cm}

Because of the van Hove singularities,
this result allows to read off the Lifshitz
exponent 

$$
\lim_{\epsilon\to 0}
\,\frac{\log(|\log(\delta\Nn(\epsilon))|)}{|\log(\epsilon)|}
\;=\;\frac{1}{2}
\mbox{ . }
$$

\noindent Moreover it 
gives precise bounds on the IDS and hence bounds on
the so-called Lifshitz constants. 
For such asymptotics to hold, it is crucial that the transfer
matrices $T^E_\sigma$ be uniformly bounded. However, it should be
possible to relax the condition
that $(\Sigma,{\bf p})$ be discrete.
By approximating a continuous measure ${\bf p}$ 
by a discrete one, a straight-forward adoption of the presented
argument allows to obtain  at least
the lower bound. Finally, the (generic) conditions 
that $E_\nu$ is only a band edge of $H_\nu$ as well as the condition
on the eigenvectors of $T^{E_\nu}_\sigma$ are not essential, but the
statement of the result would be a bit more involved.

\vspace{.2cm}

As will be discussed in Section \ref{sec-comments}, the estimates 
(\ref{eq-main}) are in a certain sense based on a 
deterministic argument, albeit taking place in a random
model. The Lifshitz constants characterizing the true behavior of the
IDS are model-dependent. One approach to calculate them is
perturbation theory. As will be sketched in 
Section \ref{sec-comments}, this also allows to go beyond the Lifshitz
tail regime.

\vspace{.2cm}

\noindent {\bf Acknowledgements:} This work was supported by the
SFB 288.

%%%%%%%%%%%%%%%%%%%%%%%%%%%%%%%%%%%%%%%%%%
\section{Setup}
\label{sec-setup}

First let us recall some of the analysis of the periodic operators
$H_\sigma$. The
eigenvalues of the transfer matrix $T_\sigma^E$ are
$\lambda=\frac{1}{2}\left(
\Tr(T_\sigma^E)+ \sqrt{\Tr(T_\sigma^E)^2-4}\right)$ and $1/\lambda$. Hence, if
$|\Tr(T_\sigma^E)|<2$, there are complex conjugate eigenvalues
$\lambda=e^{\imath \eta}$ and $e^{-\imath \eta}$
and the transfer matrix $T_\sigma^E$ is conjugate to a rotation by the phase
$\eta$. This phase is also called the rotation number and 
one speaks of the elliptic case. Then
$E\in\mbox{spec}(H_\sigma)$ and 
the IDS $\Nn_\sigma$ at $E$ is equal to $\eta/(L\pi)$ up to a
multiple of $1/L$ coming from the gap label \cite{JM,Bel}. 
On the other hand, if $|\Tr(T_\sigma^E)|>2$, the
eigenvalues are both real. One of them
has a modulus bigger than $1$ and one
smaller than $1$. This is the hyperbolic case,
$E\notin\mbox{spec}(H_\sigma)$ and the transfer
matrix is conjugate to the dilation
$\left(\begin{array}{cc} \lambda & 0 \\ 0 &
1/\lambda \end{array}\right)$. Again due to the gap labelling, one has
$\Nn_\sigma(E)=m/L$ for some positive integer $m\leq L$.
\vspace{.2cm}

Let now $E_\nu$ be a band edge of the operator $H_\nu$. 
One then has $|\Tr(T_\nu^{E_\nu})|=2$ and $\lambda=\pm 1$ 
and is therefore in the so-called
parabolic case. The transfer matrix $T_\nu^{E_\nu}$
has only one eigenvector denoted $\vec{v}$ as
well as a principal vector $\vec{w}$ satisfying $(T_\nu^{E_\nu}-\lambda
{\bf 1}) \vec{w}=\vec{v}$. Hence the basis change with
$M=(\vec{v}\;\vec{w})$ conjugates $T_\nu^{E_\nu}$ to a Jordan
normal form $M^{-1}T_\nu^{E_\nu}M=
\left(\begin{array}{cc} \lambda & 1 \\ 0 &
\lambda \end{array}\right)$. As the energy varies around
$E_\nu$, one is in
either of the above elliptic or hyperbolic cases. However, the
corresponding basis changes become singular at $E_\nu$
and it is better to work
with a basis change into an object close to the parabolic normal form.
It would be possible to simply work with the energy independent $M$, 
but for sake of more
explicit formulas later on let us choose (which is easily seen to be
possible for $|\epsilon|\leq\epsilon_0$ for some $\epsilon_0$) 
an energy dependent
basis change $M_\epsilon$ such that

\begin{equation}
\label{eq-basischange}
M_\epsilon^{-1}T_\nu^{E_\nu+\epsilon}M_\epsilon
\;=\;
\left(\begin{array}{cc} \lambda(1-\kappa_\epsilon) & 1 \\ -\kappa_\epsilon &
\lambda \end{array}\right)
\mbox{ , }
\qquad
\kappa_\epsilon
\;=\;
2-\lambda \;\Tr(T_\nu^{E_\nu+\epsilon})
\mbox{ . }
\end{equation}

\noindent As $T_\nu^{E_\nu+\epsilon}$ is a
polynomial in $\epsilon$, $M_\epsilon$ is analytic. 
If $\kappa_\epsilon<0$, one is in the hyperbolic case and
for $\kappa_\epsilon>0$ in the elliptic one. In the latter the rotation
number $\eta$ is then given by $e^{\imath
\eta}=\frac{1}{2}\left(\lambda(2-\kappa_\epsilon)+\imath 
\sqrt{4\kappa_\epsilon-\kappa_\epsilon^2}\right)$. 
Band touching happens if $\kappa_\epsilon>0$ for both positive and
negative $\epsilon$.

\vspace{.2cm}

Following \cite{JSS}, let us next define the Pr\"ufer variables
with some care. For $E\in\RR$, let $u^E$ be the sequence of real numbers 
given via the recurrence relation $H_\omega u^E=Eu^E$ and the initial
condition $u^E(-1)=\sin(\theta^0)$ and $t_\omega(0)
u^E(0)=\cos(\theta^0)$ where $t_\omega(n)=-\langle
n|H_\omega|n-1\rangle$. 
The free Pr{\"u}fer phases $\theta_\omega^{0,E}(n)$ and amplitudes
$R_\omega^{0,E}(n)>0$ are now defined by

\begin{equation}
\label{eq-freeprufer}
R_\omega^{0,E}(n) \,
\left( \begin{array}{c} \cos (\theta_\omega^{0,E}(n))
\\ \sin(\theta_\omega^{0,E}(n)) \end{array} \right)
\;=\;
\left( \begin{array}{c} t_\omega(n)
u^E(n) \\ u^E(n-1) \end{array} \right)
\mbox{ , }
\end{equation}

\noindent the above initial conditions as well as

$$
-\frac{\pi}{2}  < \theta_\omega^{0,E}(n+1) - \theta_\omega^{0,E}(n)
< \frac{3\pi}{2}
\mbox{ . }
$$

The interest will be on energies $E=E_\nu+\epsilon$ in the vicinity of
the band edge $E_\nu$, namely $|\epsilon|\leq \epsilon_0$. 
Associated to the basis change (\ref{eq-basischange}), the
$M_\epsilon$-modified Pr\"ufer variables 
$(R_\omega^{\epsilon}(n),\theta_\omega^{\epsilon}(n))\in
\RR_+\times\RR$ will be introduced next. Define a
smooth function $m_\epsilon:\RR\to\RR$ with 
$m_\epsilon(\theta+\pi)=m_\epsilon(\theta)+\pi$ and
$0 < C_1 \le  m_\epsilon' \le C_2 < \infty$, by

$$
r(\theta)\left( \begin{array}{c}
\cos (m_\epsilon(\theta)) 
\\ \sin(m_\epsilon(\theta)) \end{array} \right)
\;=\;M_\epsilon\left( \begin{array}{c}
\cos (\theta) 
\\ \sin(\theta) \end{array} \right)
\mbox{ , }
\qquad
r(\theta)>0
\mbox{ , }
\qquad
m_\epsilon(0)\in[-\pi,\pi)
\mbox{ . }
$$

\noindent Then set $\theta_\omega^{\epsilon}(n)=
m_\epsilon(\theta_\omega^{0,E_\nu+\epsilon}(n))$  and 

\begin{equation}
\label{eq-prufer}
R_\omega^{\epsilon}(n)
\left(\begin{array}{c}
\cos(\theta_\omega^{\epsilon}(n))
\\
\sin(\theta_\omega^{\epsilon}(n))
\end{array}
\right)
\;=\;
M_\epsilon \left( \begin{array}{c} t_\omega(n)\, 
u^{E_\nu+\epsilon}(n) \\ u^{E_\nu+\epsilon}(n-1)
\end{array} \right) 
\mbox{ , }
\end{equation}

\noindent where the dependence on the initial phase
$\theta_\omega^\epsilon(0)=m_\epsilon(\theta^0)$ is suppressed. 
The
oscillation theorem as proven in \cite{JSS} implies that the IDS
close to the band edge $E_\nu$ is
given by

\begin{equation}
\label{eq-IDS}
\Nn(E_\nu+\epsilon)
\;=\;
\frac{1}{\pi}\;
\lim_{n\to\infty}
\;\frac{1}{n}
\;
\EE(\theta_\omega^{\epsilon}(n))
\mbox{ , }
\end{equation}

\noindent the expectation being taken w.r.t. $\PP$. 
If ${\bf p}=\delta_\sigma$, this formula gives the IDS $\Nn_\sigma$ of
the $L$-periodic operator $H_\sigma$. A similar formula
allows to express the Lyapunov exponent in terms of the the Pr\"ufer
variables \cite{JSS}, but this will not be used here.

\vspace{.2cm}

The $M_\epsilon$-modified phase shift dynamics 
$\Ss_{\epsilon,\sigma}(\theta)$ (with energy variation $\epsilon$
relative to the band edge $E_\nu$) 
is defined via the $M_\epsilon$-modified Pr\"ufer phase with
initial condition $\theta^0=\theta$ by
$\Ss_{\epsilon,\sigma}(\theta)=
\theta^\epsilon_\omega(L)-L\,\pi\,\Nn_\sigma(E_\nu)$ where
$\omega=((\sigma_n)_{n\in\ZZ},l=0)$ and $\sigma_1=\sigma$. Note that
it verifies

\begin{equation}
\label{eq-phasedyn}
\rho\;\left(\begin{array}{c} \cos(\Ss_{\epsilon,\sigma}(\theta)) 
\\ \sin(\Ss_{\epsilon,\sigma}(\theta)) 
\end{array}\right)
\;=\;
M_\epsilon^{-1}T_\sigma^{E_\nu+\epsilon}M_\epsilon
\left(\begin{array}{c} \cos(\theta) \\ \sin(\theta) 
\end{array}\right)
\mbox{ , }
\qquad
\rho>0
\mbox{ . }
\end{equation}

\noindent One then obtains
a discrete time random dynamical system $\Ss^m_{\epsilon,\omega}$
on $\RR$ defined iteratively by: 

$$
\Ss^m_{\epsilon,\omega}(\theta)
\;=\;\Ss_{\epsilon,\sigma_m}(\Ss^{m-1}_{\epsilon,\omega}(\theta))
\mbox{ , }
\qquad
\Ss^0_{\epsilon,\omega}(\theta)\;=\;\theta
\mbox{ . }
$$

\noindent Replacing the Pr\"ufer phases in (\ref{eq-IDS})
by the phase shifts relative to the band edge $E_\nu$, the IDS
in the vicinity of $E_\nu$ is given by

\begin{equation}
\label{eq-meanrot}
\delta\Nn(\pm\epsilon)
\;=\;
\frac{1}{L}\;\frac{1}{\pi}\;
\lim_{m\to\infty}\;
\frac{1}{m}\;\EE(\Ss^m_{\epsilon,\omega}(\theta))
\mbox{ , }
\end{equation}

\noindent where the sign is chosen such that for positive $\epsilon$
one enters the spectrum of $H_\omega$.

\vspace{.2cm}

%%%%%%%%%%%%%%
\begin{figure}
\centerline{\psfig{figure=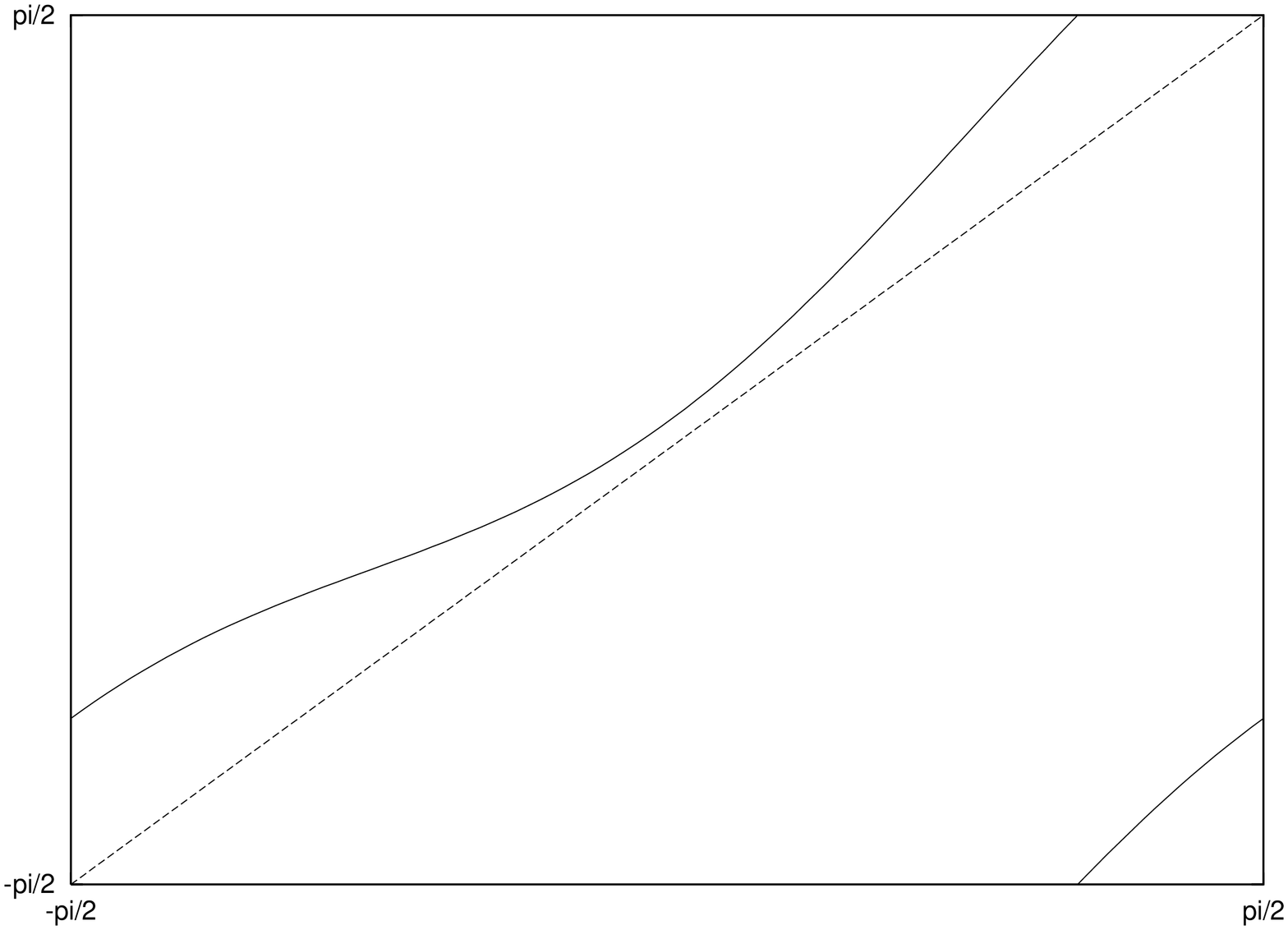,height=5cm,width=5cm,angle=270}
$\;\;\;$
\psfig{figure=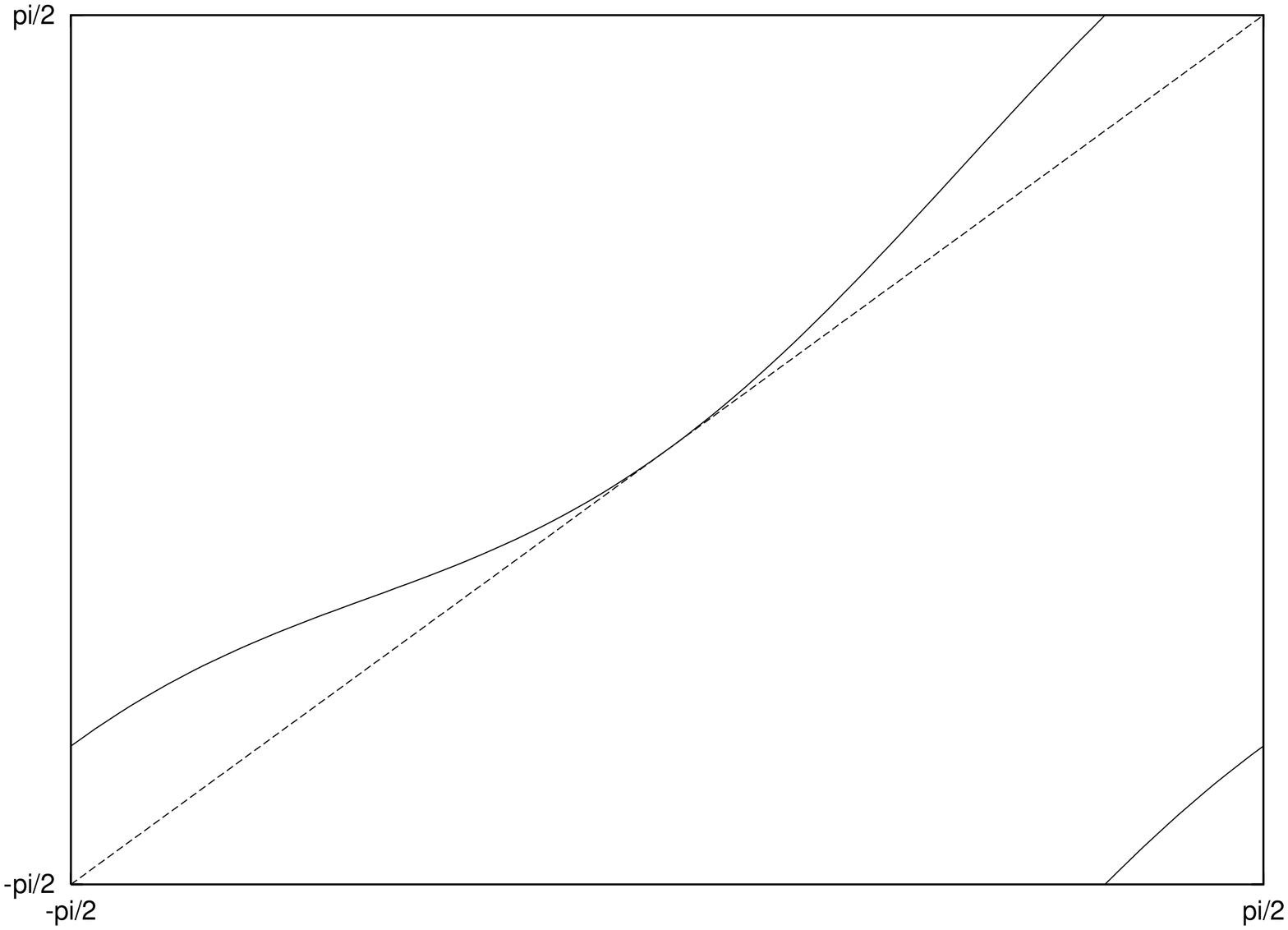,height=5cm,width=5cm,angle=270}
$\;\;\;$
\psfig{figure=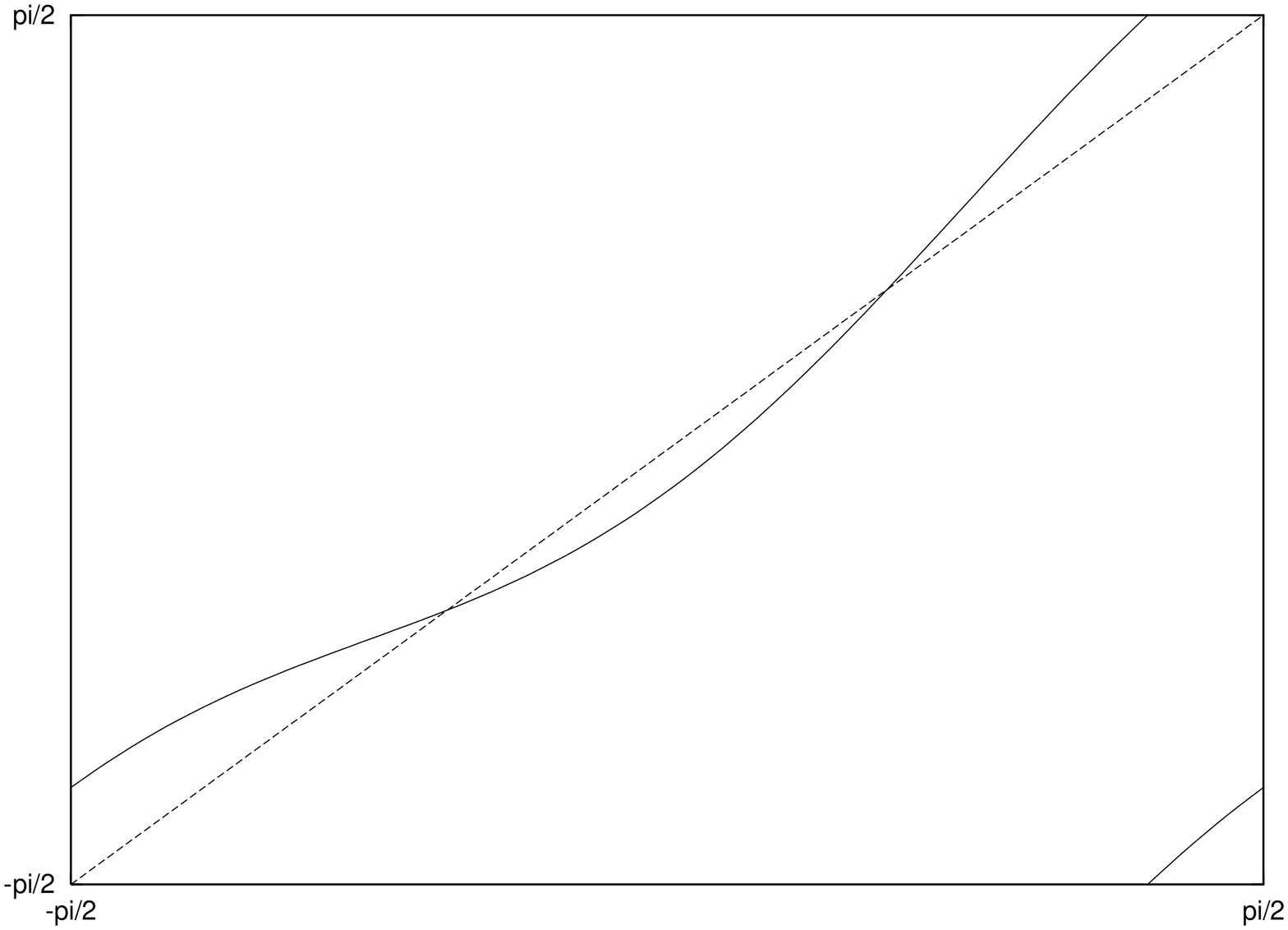,height=5cm,width=5cm,angle=270}}
\caption{{\sl Schematic plots of the phase shift dynamics 
${\Ss}_{\epsilon,\nu}$ projected on $\RR P(1)=
[-\frac{\pi}{2},\frac{\pi}{2})$ and for $\lambda=-1$ 
in the elliptic {\rm (}$\kappa_\epsilon>0${\rm )}, parabolic 
{\rm (}$\epsilon=0$ and $\kappa_\epsilon=0${\rm )}
and hyperbolic {\rm (}$\kappa_\epsilon<0${\rm )} regime.}} 
\end{figure}
%%%%%%%%%%%%%%

Comparing (\ref{eq-basischange}) and (\ref{eq-phasedyn}), one notes
that the phase shift
dynamics $\Ss_{\epsilon,\nu}$ can be immediately calculated:

$$
\tan(\Ss_{\epsilon,\nu}(\theta))
\;=\;
\frac{-\kappa_\epsilon\cos(\theta)+\lambda \sin(\theta)}{
\lambda(1-\kappa_\epsilon)\cos(\theta)+\sin(\theta)}
\mbox{ . }
$$

\noindent Alternatively the cotangent may be used. For $\lambda=-1$
the curves are plotted in Fig.~1 for three different values of
$\kappa_\epsilon$. Iterating $\Ss_{\epsilon,\nu}$ gives us a discrete
time dynamical system on $\RR$, which due to the periodicity relation
may also be regarded as the lift of a dynamical system on $\RR
P(1)\cong 
[-\frac{\pi}{2},\frac{\pi}{2})$. If $\kappa_\epsilon>0$, there is no fixed point
and therefore the dynamics $\Ss_{\epsilon,\nu}$ is conjugate to a
rotation. The rotation number can be calculated explicitely, but it is
roughly equal to $\pi$ over
the number of iterations needed to go through one period.
For $\kappa_\epsilon<0$, there are two fixed points per
period, one unstable and another one stable and (globally)
attractive so that the rotation number is $0$. In the parabolic case 
$\kappa_\epsilon=0$ there is only one fixed point, instable to one side and
stable to the other, and the rotation number is still $0$.
What was just described is locally simply a saddle node bifurcation.

\vspace{.2cm}

%%%%%%%%%%%%%%%%%%%%%%%%%%%%%%%%%%%%%%%%%%
\section{Proof}
\label{sec-proofs}

Fig.~2 shows the elliptic dynamics $\Ss_{\epsilon,\nu}$ as well as
a second dynamics which is hyperbolic. The
latter should be thought of as representing those of the hyperbolic
$\Ss_{\epsilon,\sigma}$, $\sigma\neq \nu$, which is closest to the
elliptic case. Let us first argue that the hypothesis imply that
Fig.~2 is qualitatively correct. Indeed, $E_\nu$ is supposed to be
a band edge only of
$H_\nu$ so that $\Ss_{\epsilon,\sigma}$ is hyperbolic for all
$\sigma\neq\nu$ as long as $|\epsilon|\leq \epsilon_0$ for some
adequately chosen $\epsilon_0$. Furthermore $\left(\begin{array}{c} 1
\\ 0 \end{array}\right)$ is an eigenvector of $MT^{E_\nu}_\nu M^{-1}$
which implies that the fixed point of the parabolic map $\Ss_{0,\nu}$
is $\theta=0$. For $\epsilon>0$, the map $\Ss_{\epsilon,\nu}$ is given by
shifting the graph of $\Ss_{0,\nu}$
into the elliptic regime. By hypothesis, $M^{-1}
\left(\begin{array}{c} 1\\ 0 \end{array}\right)$ is not an
eigenvector of $T^{E_\nu}_\sigma$ for any $\sigma\neq \nu$, hence the
fixed points of $\Ss_{\epsilon,\sigma}$, for $\sigma\neq \nu$ and 
$|\epsilon|\leq \epsilon_0$, are bounded away from $\theta=0$. These
facts are resumed in Fig.~2.

\vspace{.2cm}

Next let us briefly present the main argument qualitatively. According
to (\ref{eq-meanrot}), the IDS is given by the mean rotation number,
the average being taken w.r.t. the probability measure choosing the
upper and lower graph in Fig.~2 randomly. Very close to the band edge,
the dynamics $\Ss_{\epsilon,\nu}$ is only slightly in the elliptic
regime and many iterations are necessary in order to complete one
rotation. During most of these iterations, the angle is in a small
interval $I=[-a,a]$ close to the origin. If at any of these iterations
any of the other dynamics is chosen, the angle is immediately again outside
and to the left of $I$ (this will be the definition of $I$). Hence the
only way to go through $I$ and hence complete a rotation is to always 
choose the dynamics $\Ss_{\epsilon,\nu}$ until the angle is to the
right of $I$. This happens with a very small probability which leads
to the precise form of the Lifshitz tails.

%%%%%%%%%%%%%%
\begin{figure}
\centerline{\psfig{figure=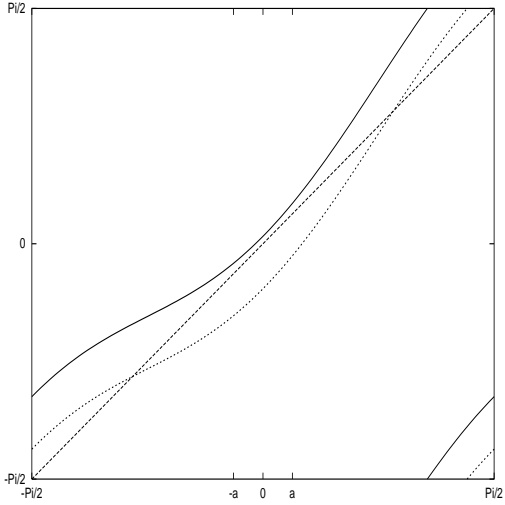,height=6.6cm,width=6.6cm,angle=270}}
\caption{{\sl Plot of the dynamics 
${\Ss}_{\epsilon,\nu}$ just in the elliptic
regime, and the next closest hyperbolic dynamics {\rm (}dotted
curve{\rm )}. This is the relevant sitution for the study of the
Lifshitz tails. 
}} 
\end{figure}
%%%%%%%%%%%%%%

\vspace{.2cm}

The proof of the lower bound now goes as follows.
The rotation number of $\Ss_{\epsilon,\nu}$ is equal to $\pi L\,
\delta\Nn_\nu(\epsilon)$. The number of iterations needed to complete
one rotation is 
$K=\left[\frac{1}{L\,\delta\Nn_\nu(\epsilon)}\right]+1$ (here
$[b]$ denotes the integer part of $b\in\RR$). Then
$\Ss^K_{\epsilon,\nu}(\theta)\geq \theta+\pi$, but 
$\Ss^{K-2}_{\epsilon,\nu}(\theta)\leq \theta+\pi$
for all $\theta\in\RR$. For sake of concreteness, suppose that $E_\nu$
is a lower band edge so that the sign in (\ref{eq-meanrot}) is $+$.
Next let us set $m=KN$ in (\ref{eq-meanrot}) and decompose
$\Ss^{KN}_{\epsilon,\omega}(\theta)$ into a telescopic sum:

\begin{equation}
\label{eq-telsum}
\delta \Nn(\epsilon)
\;=\;
\frac{1}{\pi}\,\frac{1}{L}\,\frac{1}{K}\,
\lim_{N\to\infty}\frac{1}{N} \sum_{n=1}^N
\EE\left(
\Ss^{Kn}_{\epsilon,\omega}(\theta)
-
\Ss^{K(n-1)}_{\epsilon,\omega}(\theta)
\right)
\mbox{ . }
\end{equation}

\noindent With probability $p_\nu^K$
one has $\sigma_j=\nu$ for $j=K(n-1)+1,\ldots, Kn$. In this case, one
rotation is completed and hence
each summand can be bounded from below by $\pi p_\nu^K$. Elementary
inequalities now imply the lower bound in (\ref{eq-main}).

\vspace{.2cm}

In order to prove the upper bound, set $a=\max\{\theta\leq 1\,|\,
{\Ss}_{\sigma,\epsilon}(\theta)<-\theta
\;\,\forall\;|\epsilon|\leq\epsilon_0,\;\sigma\neq\nu\}$
and $M=\inf\{m\geq
1\,|\,\Ss^m_{\epsilon,\nu}(a)\geq \pi-a \;\;\forall \;0\leq
\epsilon\leq\epsilon_0\}$. By construction, the only way to cross
$I=[-a,a]$ is to chose $\sigma=\nu$ at least
$K=\left[\frac{1}{L\,\delta\Nn_\nu(\epsilon)}\right] -M$ times. This happens
with probability $p_\nu^K$. If this event occurs, the accumulated
phase shift is of order $2a$ which can simply be bounded above by
$\pi$. Hence using the same decomposition as in (\ref{eq-telsum}), but
with the different $K$, one gets the bound $\delta\Nn(\epsilon)\leq
\frac{1}{LK}p_\nu^K$. As $M$ is finite, this implies the upper bound
in (\ref{eq-main}).

\vspace{.2cm}

%%%%%%%%%%%%%%%%%%%%%%%%%%%%%%%%%%%%%%%%%%
\section{Outlook}
\label{sec-comments}

The upper bound in the above argument exploits the following fact:
the only way 
to cross the critical region $I=[-a,a]$ is by sucessively choosing the
favorable branch $\Ss_{\epsilon,\nu}$. As this is the {\it only} way,
it seems adequate to speak of a deterministic estimate.
A more complete analysis of the mean rotation number
would also have to take into account what
happens in the remainder $S^1\backslash I$.
Obviously that heavily depends on the precise form of the maps
$\Ss_{\epsilon,\sigma}$ as well as the probability measure ${\bf p}$.

\vspace{.2cm}

A situation in which an analysis becomes feasable is perturbation
theory. Suppose that dynamics $\Ss_{\lambda,\epsilon,\sigma}$ depend
on a supplementary small parameter $\lambda$ giving the order of 
the $L^\infty$-distance between all the maps
$\Ss_{\lambda,\epsilon,\sigma}$. 
Then the random dynamics in $S^1\backslash I$ can be analysed
perturbatively in $\lambda$ allowing to calucalate the Lipshitz
constants perturbatively. This situation arises for example 
in the Anderson model
in the weak coupling limit, namely
$H_\omega=H_0+\lambda V_\omega$ with some $L$-periodic backround operator 
$H_0$ and a random potential $V_\omega$. Then the gaps of $H_0$ remain
open for sufficiently small $\lambda$ and the IDS in the gap is still
given by the gap label of $H_0$. 
Within this framework it is also
possible to study the IDS away from the Lifshitz tails. If all the
dynamics at a given energy $E$ are elliptic, then the IDS is simply given
by $\Nn(E)=\int\! d{\bf p}(\sigma)\,\Nn_\sigma(E)+\Oo(\lambda^2)$ (this
can be proven along the lines of Sec.~4.5 of \cite{JSS}). The
situation becomes more interesting at energies where there are both
elliptic and hypobolic phase shift dynamics. Then the mean rotation number
can be calculated perturbatively via a classical ruin problem
associated to the passage through an (appropriately chosen) interval
$I$. The Lifshitz tails are recovered on one extreme (where an error
leads to immediate ruin), but this
allows moreover to control a cascade of large deviation regimes
between the Lifshitz tails and the band center.
By the same techniques the Lyapunov exponent can be computed
perturbatively, completing thus the results of \cite[Thm.~14.6]{PF} and
\cite[Sec.~4.6]{JSS}. The work giving a detailed analysis corresponding to 
these ideas is under preparation.

\vspace{.2cm}

%%%%%%%%%%%%%%%%%%%%%%%%%%%%%%%%%%%%%%%%%%%

\end{document}